\documentclass[12pt]{article}
\usepackage{latexsym,amsfonts,amssymb,amsmath}
\usepackage[utf8]{inputenc}

\newcommand{\ri}{{\mathrm i}}
\newcommand{\p}{\partial}
\newcommand{\bea}{\begin{array}}
\newcommand{\eea}{\end{array}}

\catcode`@=11 \long
\def\@caption#1[#2]#3{\par\addcontentsline{\csname
ext@#1\endcsname}{#1} {\protect\numberline{\csname
the#1\endcsname}{\ignorespaces #2}} \begingroup \small
\@parboxrestore \@makecaption{\csname fnum@#1\endcsname}
{\ignorespaces #3}\par \endgroup} \catcode`@=12

\newcommand{\Q}{\mathbb{Q}}

\newcommand{\la}{\label}

\catcode`@=11 \long
\def\@caption#1[#2]#3{\par\addcontentsline{\csname
ext@#1\endcsname}{#1} {\protect\numberline{\csname
the#1\endcsname}{\ignorespaces #2}} \begingroup \small
\@parboxrestore \@makecaption{\csname fnum@#1\endcsname}
{\ignorespaces #3}\par \endgroup} \catcode`@=12

\begin{document}

\allowdisplaybreaks
 \begin{titlepage} \vskip 2cm

\begin{center} {\Large\bf Symmetries of Schr\"odinger-Pauli equations for charged  particles and quasirelativistic  Schr\"odinger equations }
\footnote{E-mail: {\tt nikitin@imath.kiev.ua} } \vskip 3cm {\bf {A.
G. Nikitin } \vskip 5pt {\sl Institute of Mathematics, National
Academy of Sciences of Ukraine,\\ 3 Tereshchenkivs'ka Street,
Kyiv-4, Ukraine, 01601\\}}\end{center} \vskip .5cm \rm

\begin{abstract}
Lie symmetries of  Schr\"odinger-Pauli equations for charged particles and quasirelativistic Schr\"odinger equations are classified. In particular a new  superintegrable system with spin-orbit coupling is discovered.
\end{abstract}
\end{titlepage}
\section{Introduction\label{int}}

Staring with the appearance of the classical works of Weil, Wigner and Wan der Warden everybody understands   the dominant role played by  symmetries in quantum mechanics. In particular all fundamental quantum mechanical systems such as Hydrogen atom and harmonic oscillator can be specified as ones having the most extended number of integrals of motion which can be admitted by Hamiltonian systems, and this property is identified as the maximal superintegrability.

Historically the systematic search for superintegrable systems started with the papers of Winternitz with collaborators  \cite{FMS}, \cite{Wint} and \cite{Wint1} where the complete description of 2d quantum mechanical systems admitting second order integrals of motion was presented. And it needs as much as twenty four years to extend this result to the case of the 3d systems, see papers \cite{evan}, \cite{evan2} and \cite{evan3}. Notice that a particular subclass of the  2d systems invariant with respect to rotations with respect to the third coordinate axis was  considered already  in \cite{Wint1}.

The second (and higher) order  symmetries  can be related to such nice properties of Schr\"odinger equation (SE) as superintegrability and supersymmetry, see surveys \cite{PW} and \cite{NNN}. We will not discuss them here but mention that searching for such symmetries is still a very popular business, and the modern trends in this field are related to the third order and even arbitrary order integrals of motion \cite{marqu, marq, snoba},  see also paper \cite{NikNik} where the determining equations for such symmetries were presented.

It was Winternitz again \cite{WY1, WY2, WY3} who starts and continued the study of superintegrable Schr\"odinger equations (SEs) generalized by the presence of the spin-orbit coupling term. The principally new feature of such equations is that they include {\it matrix} potentials with generate essential complications of the related classification procedure.

The spin-orbit coupling is a well grounded and experimentally confirmed effect which can be treated as a relativistic correction of order $\frac1{c^2}$ where $c$ is the speed of light. But there exist one more correction which is more essential since its order is $\frac1c$. This correction is described by the Pauli term which include the scalar product of the spin and magnetic fields vectors. Thus it was natural to classify superintegrable Schr\"odinger-Pauli (SP) equations.
We can mention   papers devoted to its supersymmetries \cite{N3,N4}, supersymmetries and higher order symmetries \cite{N5,N6,N7} and  Fock symmetries \cite{N8,N9,N10}.

Rather paradoxically, the less complicated  (but not less important!) Lie symmetries of the mentioned equations are still terra incognita. More exactly, everything is complete with the symmetries of the standard SE whose Lie symmetries were classified long time ago
in papers \cite{Nied}, \cite{And} and
\cite{Boy}.  Symmetries of the 3d SE with non-trivial scalar potential were described by
Boyer \cite{Boy},  whose results where corrected in \cite{Nuca}. Symmetries of this equation with scalar and vector potentials have been classified in \cite{20_2}. Symmetries of SE with a scalar (but matrix) potential are described in \cite{20_1}. Let us mention also that the complete group classification of SE equation with position dependent mass
 appears rather recently in papers \cite{NZ}  and \cite{NZ2,NN} for the stationary and time dependent
equations correspondingly. For the superintegrability aspects of such equations  see papers \cite{mill} and \cite{N1},  a superintegrable {\it relativistic} system is discussed in \cite{ang}.

In the present paper we give the completed description of all inequivalent continuous symmetries which can be accepted by time dependent SP equation for charged particles. In addition, we classify also Lie symmetries of the quasi relativistic Schr\"odinger equation which includes both Pauli, spin-orbit and Darwin  interaction terms. In this way we present a certain group-theoretical grounds for all quantum mechanical systems whose higher symmetries have been classified in the above cited papers. In particular, we give a priori information concerning symmetries which can be accepted by the related quasi relativistic models. Secondly, the related symmetry groups supply us by the equivalence relations between the systems whose higher order integrals of motion are classified \cite{Mil}. Finally, just Lie symmetries present powerful tools for construction of exact solution which sometimes have a very high value.

\section{Schr\"odinger-Pauli equation for charged particles}

Let us consider  the standard SP equation  for a charged particle with spin 1/2 interacting with the external electromagnetic field and write it in the following form:
\begin{gather}\left(\ri\frac{\p}{\p t}-H\right)\psi(t,{\bf x})=0\la{se}\end{gather}
The corresponding Hamiltonian $H$ looks as follows:
\begin{gather} \la{H1} H=\frac12\pi_a\pi_a+A^0+g \sigma_a H^a\end{gather}
where $\pi_a=p_a-eA^a, \ p_a=-\ri\frac{\p}{\p{x_a}}$, $A^0$ and $A^a$ are components of the vector-potential of the  electromagnetic field, $e$ and $g$ are coupling constants,
\begin{gather}H^a=\varepsilon_{abc}\frac{\p A^c}{\p x_b}\la{mf}\end{gather}
are components of the external magnetic field strengths, $\sigma_a$ are Pauli matrices and summation is imposed over the repeating indices $a$ over the values $a=1, 2, 3$. Moreover, the vector-potential of the external field is supposed to be time independent.

The vector-potential is supposed to satisfy the continuity equation which, in view of its time independence, is reduced to the divergenceless condition for the vector $A^a$.  However in paper \cite{20_2} we prefer to change it by the condition $A^3=0$ which always can be done using the gauge transformations. In particular  we can set:
\begin{gather}\la{vp} A^3=0,\quad A^1=\p_1\tilde F({\bf x})+\p_2G({\bf x}), \quad A^2=\p_2F({\bf x})-\p_1G({\bf x})\end{gather}
where $\p_1=\frac{\p}{\p x_1}$, etc,  $\tilde F$ and  $G$ are  functions of $\bf x$.

Alternatively using the gauge transformation $H\to \exp(-\ri  \tilde F({\bf x}))H\exp(\ri \tilde F({\bf x}))$ we can transform (\ref{vp}) to a more compact form
\begin{gather}\la{vp1} A^1=\p_2G({\bf x}), \quad A^2=-\p_1G({\bf x}),\quad A^3=F(\bf x)\end{gather}
where $F({\bf x})=\p_3\tilde F({\bf x})$. Just representation (\ref{vp1}) will be used in the present paper.

Substituting (\ref{vp1}) into (\ref{H1}) we reduce it to the following form:
\begin{gather}\la{H2}H=\frac12 p_ap_a-\frac{e}2(A^a\p_a+\p_a A^a)+V\end{gather}
where $V$ is the effective {\it matrix} potential:
\begin{gather}\la{sp}\begin{split}&V=V^0+g\sigma_aH^a,\\&
V^0=A^0+\frac{e^2}2((A^1)^2+(A^2)^2+(A^3)^2)\end{split}\end{gather}
 and
\begin{gather}H^1=\p_2F+\p_1\p_3G,\quad H^2=\p_2\p_3G-\p_1F, \quad H^3=-(\p_1^2+\p_2^2)G.\la{MFI}\end{gather}

Equation (\ref{se}) with Hamiltonian (\ref{H2}) includes three arbitrary elements, i.e., functions  $A^0$,  $F$ and $G$.

\subsection{Determining equations}

Let us search for  symmetries of the SP equation with respect
to continuous groups of transformations of dependent and independent variables. Bearing in mind the linearity of equation (\ref{se}), (\ref{H2}) we can represent the generator of the symmetry group in the form
\begin{gather}\label{so}
    Q=\xi^0\partial _t+\xi^a\partial_a+\tilde\eta\equiv \xi^0\partial _t+
    \frac12\left(\xi^a\partial_a+\partial_a\xi^a\right)+\ri\eta,
\end{gather}
where $\tilde
\eta=\frac12\xi^a_a+\ri\eta,\ \ $ $\xi^0$, $\xi^a$ and $\eta$ are
functions of $t,{\mathbf x} $  and $\p_t=\frac{\p}{\p t}$. Moreover, $\eta$ is a $2\times 2$ matrix which can be expanded via Pauli matrices:
\begin{gather}\la{eta}\eta= \eta^0\sigma_0+\sigma_a\eta^a\end{gather}
where $\sigma_0$ is the $2\times2$ unit matrix which will be omitted in the following formulae.
In contrary, $\xi^0$ and $\xi^a$ are scalar functions which are multipliers for the unit matrix.

Generator (\ref{so}) transforms solutions of equation (\ref{se}), (\ref{H2})  into solutions provided  the following operator equation is satisfied:
\begin{gather}\la{ic}[Q,L]\equiv QL-LQ=\alpha L\end{gather}
where $L=\ri\p_t-H$ and $\alpha$ is an unknown function of $t$ and $\bf x$.

Evaluating the commutator in the l.h.s. of (\ref{ic}) and equating coefficients for the linearly independent
differentials we obtain the following system of  equations for unknowns $\xi^0, \xi^a, \eta,\  V $ and $\alpha$:
\begin{gather}
\dot\xi^0=-\alpha, \quad  \xi^0_a=0,\la{de1}\\
\la{de2} \xi^b_{a}+\xi^a_{b}-\frac{2}3\delta_{ab}\xi^i_i=0,
\\\label{de6} 2\xi^i_i=-3\alpha,\\ \dot\xi^a+\eta_a
=e(A^b\xi^a_b-\xi^bA^a_b+\alpha A^a),\label{de7}\\
\label{de8}\xi^aV_a=\alpha V+\dot\eta+\ri[\eta,V]-eA^a\eta_a\end{gather}
where $\dot\eta=\frac{\p \eta}{\p t} $ and $\eta_a=\frac{\p \eta}{\p x_a}$.

Thus to find symmetries of SP equation we are supposed to solve the system of  equations (\ref{de1})--(\ref{de8}) which is rather complicated. In particular this system includes two matrix equations (\ref{de7}) and  (\ref{de8}).  Collecting in (\ref{de7}) the terms proportional to the unit matrix we obtain:
\begin{gather}\la{con2} \dot\xi^a+\eta^0_a
=e(A^b\xi^a_b-\xi^bA^a_b+\alpha A^a)\end{gather}
while the remaining terms are reduced to the following condition:
\begin{gather}\la{con1}\frac{\p{\eta^a}}{\p x_b}=0.\end{gather}

Equation (\ref{de8}) also is decoupled to the scalar and vector parts:
\begin{gather}\label{con3}\xi^aV^0_a=\alpha V^0+\dot\eta^0-eA^a\eta^0_a,\end{gather} and
\begin{gather}\label{con4}g(\xi^aH_a^b-\alpha H^b+2\varepsilon^{bcd}\eta^cH^d)-\dot\eta^b=0\end{gather}
where $\varepsilon^{bcd}$ is the absolutely antisymmetric unit tensor.

Substituting (\ref{con2}) into (\ref{con3})  and using definition (\ref{sp})  we reduce the latter equation to the following form:
\begin{gather}\la{con8}\xi^aA^0_a+\dot\xi^aA^a=\alpha A^0+\dot \eta^0. \end{gather}

In other words, system (\ref{de1})--(\ref{de8}) includes the autonomous subsystem  formed by equations (\ref{de1}), (\ref{de2}), (\ref{de6}), (\ref{con2}) and (\ref{con3}). Its solutions, which give symmetries of equation (\ref{se}), (\ref{H2}) in the particular case $g=0$, have been already obtained in paper \cite{20_2}. On the other hand setting $A^1=A^2=A^3=0$ we reduce our classification problem to description of symmetries of the SP equation for chargeless  particles. This description had been made in paper \cite{20_1}. The more general problem with nontrivial coupling constant $g$ and nontrivial vector potentials is solved in the present paper.

\subsection{Classification of symmetries}
In accordance with the above the first necessary step in description of symmetries of the SP equation for charged  particles is the group classification of the standard SEs including the vector potential of the external electromagnetic field. Then going over all inequivalent versions of such external fields found in paper \cite{20_2} we will select such of them which are compatible with equation (\ref{con4}) .
By construction equations (\ref{se}), (\ref{H1}) with arbitrary potentials admit two dimensional symmetry algebra spanned on $P_0=\ri \frac{\p}{\p t}$ and the unit operator.
In accordance with the results presented in \cite{20_2} there exist six inequivalent cases when this algebra is extended to three dimensional one. The corresponding potentials  and the related additional symmetries $Q$ are given by the following formulae:
\begin{gather}\la{2}\begin{split}&A^1= \p_2G({ x_1, x_2}),\quad A^2= - \p_1G({x_1,x_2}),\\&A^3=F(x_1,x_2),\quad A^0=R(x_1,x_2),\end{split}\\\la{2_1}P_3=p_3=-\ri \p_3\end{gather}
where $F(.,.),\ G(.,.)$ and $R(.,.)$ are arbitrary functions of the arguments fixed in the brackets;
\begin{gather}\la{3}\begin{array}{l}A^1=\p_2G(\tilde r,x_3) \quad A^2=-\p_1G(\tilde r,x_3),\\A^3=F(\tilde r,x_3),\quad A^0=R(\tilde r,x_3)+\kappa\varphi,\end{array} \\\la{3_1} Q_1=L_3+\kappa t\end{gather}
where $\tilde r=\sqrt{x_1^2+x_2^2}, \ \varphi=\arctan\frac{x_2}{x_1},  L_3=x_1p_2-x_2p_1$ and $\kappa$ is an arbitrary real parameter;
\begin{gather}\la{4}\begin{array}{l}A^1=\p_2 G(\tilde r, \varkappa),\quad A^2=-\p_1G(\tilde r, \varkappa),\\A^3=F(\tilde r,\varkappa),\quad A^0=R(\tilde r,\varkappa)+\kappa \varphi,\end{array}\\ Q_2=L_3+p_3+\kappa t\la{4_1}\end{gather}
where $ \varkappa=\varphi-x_3$;
\begin{gather}\la{5}\begin{array}{l} A^1=\p_2G(\theta,\mu),\quad A^2=-\ln(r)\p_1G(\theta,{\mu})\\A^3=\ln(r)( F(\theta,{\mu})+\tilde F(\nu,\mu)),\quad A^0=\frac1{r^2} R(\theta,{\mu}),
\end{array}\\Q_3=D+\alpha L_3,\la{5_1}
\end{gather}
where $\mu=\alpha\rho{-\varphi},\ \nu=\rho+\alpha\varphi,\ \rho=\ln(\tilde r)$ and
$D=2tP_0-x_ap_a+{3\ri }/2;$
\begin{gather}\la{6}\begin{array}{l}A^1= \p_2G({ x_1, x_2}),\quad  A^2= - \p_1G({x_1,x_2}),\quad A^3=F({x_1,x_2}),\\ A^0= R({x_1,x_2})-\frac{\omega^2x_3^2}2+\omega x_3 F(x_1,x_2),\end{array}\\B^3=\exp( \omega t)(p_3-\omega x_3);\la{6_1}
\end{gather}
\begin{gather}\la{7}\begin{array}{l}A^1=\p_2G(\theta,\varphi),\quad A^2=-\p_1G(\theta,\varphi),\quad A^3=\frac1r(\tilde F(\theta,\varphi)+(\p_\rho+2)F(\rho,\varphi)) ,\\A^0=-\frac{\omega^2 r^2}2+\frac1{r^2}R(\theta,\varphi)+4\omega \p_\rho F(\rho,\varphi), \end{array}\\A^+=\exp(2\omega t)(p_0+\omega^2r^2-
\frac{\omega}2(x_ap_a+p_ax_a)).\la{7_1}\end{gather}

Notice that $L_3$ is nothing but the third component of the orbital moment. Up to rotation it can be replaced by the first component $L_1=x_2p_3-x_3p_2$ or the second component $L_2=x_3p_1-x_1p_3$.

Consider consequently all potentials and symmetries (\ref{2})--(\ref{7_1})  in more detail and verify their compatibility with equation (\ref{con4}). The vector of magnetic field corresponding to potentials (\ref{2}) has the following components (refer to (\ref{MFI})):
\begin{gather}\la{8}H^1=-\p_2F({ x_1, x_2}),\quad H^2=\p_1F({ x_1, x_2}),\quad H^3=-(\p_1^2+\p_2^2)G({ x_1, x_2}).\end{gather}

The related symmetry  (\ref{2_1}) includes only a part of matrix  $\eta$ which is proportional to the unit matrix, i.e., only the first term from the r.h.s. of equation (\ref{eta}). Whenever the Pauli term is present, it is necessary to add the remaining part of $\eta$ equal to $\sigma^a\eta^a$, i.e., replace (\ref{2_1}) by the following operator:
\begin{gather}\la{2_2}\hat P_3=p_3+\eta^0+\eta^a\sigma^a\end{gather}
where $\eta^0$  and  $\eta^a$ are unknown functions.
 The corresponding  condition (\ref{con4}) is reduced to the following form:
\begin{gather}\label{9}g(\p_3H^b+2\varepsilon^{bcd}\eta^cH^d)-\dot\eta^b=0.\end{gather}

All components of the magnetic field presented in (\ref{8}) do not depend on $x_3$, the same is true for functions $\eta^a$ in view of equation (\ref{con1}). This means that condition (\ref{9})  nullifies all functions $\eta^a, a=1, 2, 3$ if $F({ x_1, x_2})$ and $G({ x_1, x_2})$ are arbitrary. However, for the very special case of these functions corresponding to {\it constant} magnetic field (\ref{8}) equation (\ref{9}) has non-trivial solutions which, up to rotation transformations, can be represented in the following form:
\begin{gather}\la{10}\begin{split}& H^1=H^2=0, \ H^3=c_1, \\&
\eta^1=c_2\sin(2c_1g t)+c_3\cos(2c_1g t),\\&\eta^2=c_2\cos(2c_1g t) - c_3 \sin (2c_1g t), \eta^3=c_4 \end{split}\end{gather}
where $c_1,...,c_4$ are arbitrary parameters. Moreover, up to scaling of the independent variables and the routine  trigonometric manipulations we can reduce these parameters to the form $c_1=1, c_2=0,\ c_3=1, c_4=c$.

The next system we consider is given by relations (\ref{3}). The corresponding symmetry (\ref{3_1}) again has to be generalised by adding matrix $\eta$ and written in the form:
\begin{gather}\la{3_2} Q=L_3+\eta^0+\sigma^a\eta^a.\end{gather}

The components $\eta^a$  should satisfy condition (\ref{con4}) which is reduced to the following form:
\begin{gather}\la{11}g((x_1\p_2-x_2\p_1)H^b+2\varepsilon^{bcd}\eta^cH^d)-\dot\eta^b=0\end{gather}
or
\begin{gather}\la{12}g(\p_\varphi H^b+2\varepsilon^{bcd}\eta^cH^d)-\dot\eta^b=0.\end{gather}

But in accordance with (\ref{MFI}) and (\ref{3}) the related components of the magnetic field can be represented as:
\begin{gather}\la{13}H^1=-\p_3(x_2G^1-x_1G^2),\quad H^2=\p_3(x_1G^1+x_2G^2),\quad H^3=(2+\tilde r \p_{\tilde r})G^2\end{gather}
where $G^1$ and $G^2$ are functions of $\tilde r$ and $x_3$. In view of  (\ref{13}) equations  (\ref{12}) are reduced to the following form:
\begin{gather}\la{14}2g(\eta^1H^2-\eta^2H^1)=\dot\eta^3,\\\begin{split}&g(2\eta^2H^3+(2\eta_3-1)H^1)=\dot\eta^1,
\\& g(2\eta^1H^3+(1-2\eta_3)H^2)=-\dot\eta^2.\la{15}\end{split}\end{gather}

For generic magnetic field whose components are given by equation (\ref{13}) we obtain from (\ref{14}) that
\begin{gather}\la{15a}\eta^1=0,\quad \eta^2=0,\quad \eta^3=\frac12.\end{gather}
However, for the constant magnetic field there is one more solution given by equations (\ref{10}).

In the complete analogy with the above it is possible to consider the remaining systems (\ref{4})--(\ref{7}) and prove, that all symmetries (\ref{4_1}), (\ref{5_1}), (\ref{6_1}) and (\ref{7_1}) valid for the Schr\"odinger equation are kept for the Schr\"odinger-Pauli  equation
also, provided $L_3$ is changed to $J_3=L_3+\frac12\sigma_3$. The main point of the proof is that $H^3$ always depends on invariant variables for these symmetries, and so in all cases we have the same equation (\ref{15}) which nullifies either $\eta^1, \eta^2$ or $H^1, H^2$.

The obtained results can be formulated as the following statement.

Theorem. {\it All Lie symmetries of the Schr\"odinger equation with time independent scalar and vector potentials can be extended to the symmetries of  the corresponding SP equation, if we change the orbital momentum  operators $L_a$ by the total orbital momentum  operators $J_a=L_a+\frac 12\sigma_a$ . In addition, for constant magnetic fields or for the fields with the only nontrivial component $H^3$, the additional matrix symmetries }
\begin{gather}\la{23} Q=\sigma_1\cos(2gt) +\sigma_2\sin(2gt)+c\sigma_3
 \quad
{\it and } \quad \tilde Q=\sigma_3
\end{gather}
{\it are valid}.

Since the symmetries of the Schr\"odinger equation with time independent scalar and vector potentials have been described in paper  \cite{20_2}, in fact we have in hands all inequivalent  symmetries of the related SP equation. In the following section we represent them together with the corresponding arbitrary elements $F, G$ and $A^0$.

\subsection{Classification results}

Using Theorem 1 and the results presented in paper  \cite{20_2} we can effectively describe all inequivalent Lie symmetries of the SP equation. Let us present them in the following  tables. In accordance with  \cite{20_2} the dependence of such symmetries on time can be linear, quadratic or exponential. In additional, these symmetries can be time independent.

 In the tables  $J_a=L_a+\frac12\sigma_a, a=1,2,3, $ and
\begin{gather}\la{gen} \begin{split}& L_a=\varepsilon_{abc}x_bp_c,\\ &P_a=p_a,\ G_a=tP_a-x_a\end{split}\end{gather}
\begin{gather}\la{agent}\begin{split}& D=2tP_0-x_ap_a+{3\ri }/2,\\& A=tD-t^2P_0+{r^2}/2\end{split}\end{gather} are symmetries of the free SE. In addition,  \begin{gather}\la{opop}\begin{split}&A^{+}(\omega)=\exp(2\omega t)(P_0+\omega^2r^2-
\frac{\omega}2(x_aP_a+P_ax_a)),
\\&B^{\pm}_a(\omega)=
\exp(\pm \omega t)(P_a\mp\omega x_a), a=1, 2, 3\end{split}\end{gather} are symmetries of SE with repulsive potential. In addition, $F(.), G(.), R(.)$ and $\tilde R(.)$ are arbitrary functions of the arguments fixed in the brackets, and the Greek letters denote arbitrary coefficients.

In contrast with \cite{20_2}, in the classification tables  we will specify inequivalent arbitrary elements $F, G$ and $A^0$ instead of direct presentation of vector potentials. In this way we obtain the tables in a rather compact form. In addition, we arrange the classification results in the  manner convenient for presentation of symmetries of  more general equation including spin-orbit and Darwin terms. Just these equations are discussed in the following section.

\newpage

\begin{center}Table 1.
Generating functions and symmetries linear in $t$
\end{center}
\begin{tabular}{l l l l l l}
\hline No&\hspace{-3mm}Functions $F$&\hspace{-3mm}Functions $G$&\hspace{-3mm}Potential $A^0$&Symmetries&\hspace{-3mm}Potential $S$
\vspace{2mm}

\\
\hline

\vspace{2mm}

1. &\hspace{-3mm}$  F(x_1,x_2)$&\hspace{-3mm}$ G(x_1,x_2)$&\hspace{-3mm}$R_0(x_1,x_2)$&\hspace{-3mm}$P_3$&\hspace{-3mm}$\tilde R_0(x_1,x_2)$\\
\vspace{2mm}
2.&\hspace{-3mm} $F(\tilde r,x_3)$&\hspace{-3mm}$G(\tilde r, x_3)$&\hspace{-3mm}$R(\tilde r, x_3)+\kappa\varphi$&\hspace{-3mm}$J_3+\kappa t$&\hspace{-3mm}$\tilde R(\tilde r, x_3)$\\
\vspace{1mm}
3.&\hspace{-3mm}$F(\tilde r, \varphi-x_3)$&\hspace{-3mm}$G(\tilde r, \varphi-x_3)$&\hspace{-3mm}$\begin{array}{l}R(\tilde r, \varphi-x_3)\\+\kappa \varphi\end{array}$&\hspace{-3mm}$J_3+\kappa t+P_3$&\hspace{-3mm}$\tilde R(\tilde r, \varphi-x_3)$\\
\vspace{1mm}

4. &\hspace{-2mm}$\tilde F(\tilde r)$&\hspace{-3mm}$G(\tilde r)$&\hspace{-3mm}$R(\tilde r)+\kappa \varphi$&\hspace{-3mm}$P_3,\  J_3+\kappa t$&\hspace{-3mm}$R(\tilde r)$\\

\vspace{1mm}

5.&\hspace{0mm}$0$&\hspace{-3mm}$G(\tilde r)$ &\hspace{-3mm}$R(\tilde r)+\kappa\varphi$&\hspace{-3mm}
$P_3,\ J_3+\kappa t,  \ G_3$&\hspace{-3mm}$\tilde R(\tilde r)$ \\

\vspace{1mm}

6.&\hspace{0mm}$0$&\hspace{-3mm}$G(x_1,x_2)$ &\hspace{-3mm}$R(x_1,x_2)$&\hspace{-3mm}$P_3,\ G_3$&\hspace{-3mm}$\tilde R(x_1,x_2)$\\

\vspace{1mm}

7.&\hspace{-3mm}$x_1F(x_3)$&\hspace{-3mm}$x_1G(x_3)$&\hspace{-3mm}$R(x_3)$&\hspace{-3mm}$P_1, \ P_2$&\hspace{-3mm}$\tilde R(x_3)$\\

\vspace{1mm}

8.&\hspace{0mm}$0$&\hspace{-3mm}$G(\tilde r)$&\hspace{-3mm}$R(\tilde r)$&\hspace{-3mm}$J_3,\ P_3,
G_3$&\hspace{-3mm}$\tilde R(\tilde r)$\\

\vspace{2mm}

9.&\hspace{0mm}$0$&\hspace{-3mm}$x_2\Phi(x_3)$&\hspace{-3mm}$R(x_3)$&\hspace{-3mm}$P_1,\ P_2, \ G_2 $&\hspace{-3mm}$\tilde R(x_3)$\\

\vspace{2mm}

10.&\hspace{0mm}$0$&\hspace{0mm}$x_3\varphi/r$&\hspace{-3mm}$R(r)$&\hspace{-5mm}
$\begin{array}{l}J_1+\lambda (x_3^2\p_2\varphi -x_1)/r,\\
J_2-\lambda ( x_3^2\p_1\varphi- x_2)/r, \\ J_3\end{array}$&\hspace{-3mm}$\tilde R(r)$\\

\vspace{2mm}

11.&\hspace{-3mm}$\begin{array}{l}x_1F(x_3)+\alpha x_1 x_2\end{array}$&\hspace{-3mm}$\begin{array}{l}x_1G(x_3)+\alpha x_1^2\end{array}$&\hspace{-3mm}$R(x_3)
$&\hspace{-5mm}$\begin{array}{l}P_1+\alpha x_2, P_2-\alpha x_1\end{array}$&\hspace{-3mm}$\tilde R(x_3)$\\

\vspace{1mm}

12.&\hspace{-3mm}$\begin{array}{l}\nu x_1\cos(x_3)\\+\alpha x_1 x_2\end{array}
$&\hspace{-3mm}$\begin{array}{l}\alpha x_1^2\\-\nu x_1\sin(x_3)\end{array}$&\hspace{-3mm}$R(x_3)$&\hspace{-5mm}$\begin{array}{l}P_1+\alpha x_2,\\ P_2-\alpha x_1, J_3+P_3\end{array}$&\hspace{-3mm}$\tilde R(x_3)$\\

\vspace{1mm}

13.&\hspace{-3mm}$\alpha x_1 x_2$&\hspace{-3mm}$\alpha x_1^2$&\hspace{-3mm}$R(x_3)$&\hspace{-5mm}$\begin{array}{l}P_1+\alpha x_2, Q, \tilde Q\\ P_2-\alpha x_1,\ J_3\end{array}$&\hspace{-3mm}$\tilde R(x_3)$\\

\vspace{2mm}

14.&\hspace{-3mm}$\alpha x_1 x_2$&\hspace{-3mm}$\alpha x_1^2$&\hspace{0mm}0&\hspace{-3mm}$\begin{array}{l}P_1+\alpha x_2, P_2-\alpha x_1,\\ P_3, G_3, J_3,  Q,  \tilde Q\end{array}$&\hspace{0mm}0
\vspace{1mm}

\\
\hline\hline

\end{tabular}
\begin{center}Table 2.
Generating functions and symmetries including dilatation
\end{center}
\begin{tabular}{l l l l l l}
\hline No&Functions $F$&\hspace{-3mm}Functions $G$&Potential $A^0$&Symmetries&Potential $ \tilde A$
\vspace{2mm}

\\
\hline

\vspace{2mm}

1. &\hspace{-5mm}$\begin{array}{l}F(r^{2\alpha},r^\alpha\text{e}^{-\varphi})\\+\tilde F(\theta,r^\alpha\text{e}^{-\varphi})
\end{array}$&\hspace{-3mm}$G(\theta,r^\alpha\text{e}^{-\varphi})$&
$ \frac1{r^2}{R(\theta,r^\alpha\text{e}^{-\varphi})}$&$\begin{array}{l} D+\alpha J_3 \end{array}$&$\ln(r){\tilde R(\theta,r^\alpha\text{e}^{-\varphi})}$
\vspace{1mm}\\
2.&\hspace{-3mm}$\frac1{\tilde r} \tilde F(\varphi)$&\hspace{-3mm}$G(\varphi)$&$\frac1{\tilde r^2}R(\varphi)$&$D,
P_3$&$\ln(r){\tilde R(\theta,r^\alpha\text{e}^{-\varphi})}$\\
\vspace{1mm}
3.&\hspace{-3mm}$F(\varphi,\theta)$&\hspace{-3mm}$G(\varphi,\theta)$&$\frac1{r^2}R(\varphi,\theta)$&$A, \ D$&$\ln(r){\tilde R(\theta,r^\alpha\text{e}^{-\varphi})}$\\
\vspace{1mm}

4.&\hspace{-3mm}$0$&$\frac{\lambda}{r} x_3\varphi$&$\frac\mu{r^2}$&\hspace{-2mm}$\begin{array}{l}\frac\lambda{r} (x_3^2\p_2\varphi- x_1)+J_1,\\
\frac\lambda{r} (x_2 +x_3^2\p_1\varphi )-J_2, \\ D, \ A, J_3\end{array}$&$\mu\ln({r})$\\

5.&\hspace{-3mm}$F(\theta)$&\hspace{-3mm}$G(\theta)$&$\frac1{r^2}R(\theta)$&$A, \ D, \ J_3$&$\ln(r)\tilde R(\theta)$
\vspace{1mm}\\

6.&\hspace{-3mm}$0$&\hspace{-3mm}$G\left(\tilde r^\alpha\text{e}^{-\varphi}\right)$&$\frac1{r^2}R\left(\tilde r^\alpha\text{e}^{-\varphi}\right)$&\hspace{-2mm}$\begin{array}{l}D+\alpha L_3,\\
P_3,\ G_3\end{array}$&$\ln(r)\tilde R\left(\tilde r^\alpha\text{e}^{-\varphi}\right)$\\

\vspace{2mm}

7.&\hspace{-3mm}$F(\varphi)$&\hspace{-3mm}$G(\varphi)$&$\frac1{r^2}R(\varphi)$&$\begin{array}{l}A, \ D, \\ P_3, \ G_3\end{array}$&$\ln({r})\tilde R(\varphi)$\\
\vspace{1mm}
8.&$0$&\hspace{-3mm}$\lambda\frac{ x_2}{x_3}$&$\frac\mu{x_3^2}$&$\begin{array}{l}P_1,\ P_2, \ G_2 ,\\ D, \ A \end{array}$&$\kappa\ln({x_3})$

\vspace{2mm}\\

\hline\hline
\end{tabular}

 \vspace{2mm}

\newpage

\begin{center}Table 3.
Generating functions and  symmetries exponential in $t$.
\end{center}
\begin{tabular}{l l l l l l}
\hline No&\hspace{-3mm}Function $F$&\hspace{-3mm}Function $G$&Potential $A^0$&Symmetries&Potentials $\tilde A$
\vspace{1mm}
\\
\hline
\vspace{1mm}

1.&\hspace{0mm}$0$&\hspace{-3mm}$G(x_1,x_2)$ &$R(x_1,x_2)-\frac{\omega^2x_3^2}2$&\hspace{-3mm}$\begin{array}{l}B^+_3(\omega) , B^-_3(\omega) \end{array}$&$\tilde R(x_1,x_2)$\\
\vspace{1mm}

2.&\hspace{0mm}$0$&\hspace{-3mm}$G(\tilde r)$ &\hspace{-3mm}$R(\tilde r)-\frac{\omega^2x_3^2}2$&\hspace{-3mm}$\begin{array}{l}B^+_3(\omega) , B^-_3(\omega), \ J_3\end{array}$&$\tilde R(\tilde r)$\\
\vspace{1mm}

3.&\hspace{0mm}$0$&\hspace{-3mm}$G(x_1)$ &\hspace{-3mm}$R(x_1)-\frac{\omega^2x_3^2}2$&\hspace{-3mm}$\begin{array}{l}B^+_3(\omega), B^-_3(\omega), \\ P_2,\ G_2\end{array}$&$\tilde R(x_1)$\\
\vspace{1mm}

4.&\hspace{-3mm} $x_1F(x_3)$&$\hspace{-3mm}x_1G(x_3)$&\hspace{-3mm}$\begin{array}{l}R(x_3)\\+\omega_1x_1F(x_3)\\+\omega_2x_2G(x_3)\\-\frac{\omega_1^2x_1^2}2-
\frac{\omega_2^2x_2^2}2
\end{array}$&\hspace{-3mm}$\begin{array}{l} B^+_1(\omega_1),  B^-_1(\omega_1),\\ B^+_2(\omega_2), B^-_2(\omega_2) \end{array}$&$\tilde R(x_3)$\\

\vspace{1mm}5.&\hspace{-3mm}$F(x_1,x_2)$&\hspace{-3mm}$G(x_1,x_2)$&\hspace{-3mm}$
\begin{array}{l}R({x_1,x_2})-\frac{\omega^2x_3^2}2\\+\omega x_3 F(x_1,x_2)\end{array}$&\hspace{-2mm}$B^+_3(\omega)$&$\tilde R(x_1,x_2)$\\\vspace{1mm}

6.&\hspace{0mm}$0$&\hspace{-3mm}$G(\tilde r)$&\hspace{-3mm}$R({\tilde r})-\frac{\omega^2x_3^2}2+\kappa \varphi$&\hspace{-3mm}$\begin{array}{l}B^+_3(\omega),  B^-_3(\omega)
,\\ L_3+\kappa t \end{array}$&$\tilde R({\tilde r})$\\

\vspace{1mm}

7.&\hspace{0mm}$0$&\hspace{-3mm}$G(\tilde r)$&\hspace{-3mm}$\begin{array}{l}R({\tilde r})-\frac{\omega^2x_3^2}2\\+\omega x_3 F(\tilde r)+\kappa \varphi\end{array}$&\hspace{-3mm}$\begin{array}{l}B^+_3(\omega)
,\\ L_3+\kappa t \end{array}  $&$\tilde R({\tilde r})$\\

\vspace{1mm}
8.&\hspace{-3mm}$F(x_1)$&\hspace{-3mm}$G(x_1)$&\hspace{-3mm}$\begin{array}{l}R({x_1})-\frac{\omega^2x_3^2}2\\+\omega x_3 F(x_1)\end{array}$&\hspace{-3mm}$\begin{array}{l}B^+_3(\omega),   P_2, \text{and }\\ G_2\text{ if }G(x_1)=0
\end{array} $&$\tilde R({x_1})$\\
\vspace{1mm}

9.&\hspace{-3mm}$\begin{array}{l}\frac1rF(\theta,\varphi)\end{array}$&\hspace{-3mm}$G(\theta,\varphi)$
&\hspace{-3mm}$ \begin{array}{l}
\frac1{r^2}R(\theta,\varphi)-\frac{\omega^2 r^2}2\\+\omega\p_\varphi G(\theta,\varphi)\\+\omega\cos(\theta) F(\theta,\varphi)\end{array}$&$A^{+}(\omega)$&$\ln(r)\tilde R(\theta,\varphi)$\\
\vspace{1mm}
10.&\hspace{-3mm}\vspace{1mm}$\begin{array}{l}\frac1{\tilde r}F(\varphi)\end{array}$&\hspace{-3mm}$G(\varphi)$&\hspace{-3mm}$ \begin{array}{l}
 \frac1{r^2}R(\theta,\varphi)-\frac{\omega^2r^2}2\\+\frac{\omega x_3 F(\varphi)}{\tilde r}\end{array}$&\hspace{-3mm}$\begin{array}{l}B^+_3(\omega),  A^{+}(\omega)\end{array}$&$\ln(r)\tilde R(\theta,\varphi)$\\
\vspace{1mm}
11.&\hspace{-3mm}$\begin{array}{l}\frac1rF(\theta)\end{array}$&\hspace{-3mm}$G(\theta)$&\hspace{-3mm}$ \begin{array}{l}
\frac1{r^2}R(\theta)-\frac{\omega^2 r^2}2\\+\omega\cos(\theta)F(\theta)\end{array}$&\hspace{-3mm}$\begin{array}{l}A^{+}(\omega), J_3\end{array}$&$\ln(r)\tilde R(\theta)$\\
\vspace{1mm}
12.&\hspace{-3mm}$\begin{array}{l}\frac{1}{\tilde r}\end{array}$&$0$&\hspace{-3mm}$ \begin{array}{l}
\frac1{\tilde r^2}R(\varphi)\\-\frac{\omega^2r^2}2+\frac{\omega x_3}{\tilde r}\end{array}$&\hspace{-3mm}$\begin{array}{l}A^{+}(\omega), B^+_3(\omega) \end{array}
$&$\ln({\tilde r})\tilde R(\varphi)$\\

13.&\hspace{-3mm}$\begin{array}{l}\frac{\nu}{x_1}\end{array}$&0&\hspace{-3mm}$-\frac{\omega^2r^2}2
+\frac{\omega x_3}{x_1} $&\hspace{-3mm}$\begin{array}{l}A^{+}(\omega), B_2^+, B_2^-, B^+_3(\omega) \end{array}
$&\hspace{3mm}$0$\vspace{1mm}
\\
14.&\hspace{-3mm}$\begin{array}{l}x_1F(x_3)\\+\alpha x_1 x_2\end{array}$&$\begin{array}{l}x_1G(x_3)\\+\alpha x_1^2\end{array}$&\hspace{-3mm}$\begin{array}{l}R(x_3)-\frac{\omega^2\tilde r^2}2\\-2\alpha\omega x_1x_2\end{array}$&\hspace{-3mm}$\begin{array}{l} B^+_1(\omega)+\alpha \text{e}^{\omega t}x_2,\\ B^-_2(\omega)-\alpha \text{e}^{-\omega  t}x_1\end{array}$&$\tilde R(x_3)$\\
\vspace{1mm}

15.&\hspace{-3mm}$\alpha x_1 x_2$&$\alpha x_1^2$&$-\frac{\omega^2x_3^2}2$&\hspace{-3mm}$\begin{array}{l} P_1+\alpha x_2,  P_2-\alpha x_1,\\ B^+_3(\omega),  B^-_3(\omega), L_3,\\Q,\  \tilde Q
\end{array}$&\hspace{3mm}$0$\\
\vspace{1mm}

16.&\hspace{-3mm}$\alpha x_1 x_2$&$\alpha x_1^2$&\hspace{-3mm}$\begin{array}{l}-2\alpha\omega x_1x_2\\ -\frac{\omega^2\tilde r^2}2\end{array}$&\hspace{-3mm}$\begin{array}{l} B^+_1(\omega)+\alpha \text{e}^{\omega t}x_2,\\ B^-_2(\omega)-\alpha \text{e}^{-\omega  t}x_1,\\P_3,  G_3 , Q,  \tilde Q \end{array}$&\hspace{3mm}$0$\\
\vspace{1mm}

17.&\hspace{-3mm}$\alpha x_1 x_2$&\hspace{-3mm}$\alpha x_1^2$&\hspace{-3mm}$\begin{array}{l}-2\alpha\omega x_1x_2\\-\frac{\omega^2\tilde r^2}2-\frac{\omega_3^2x_3^2}2\end{array}$&\hspace{-3mm}$\begin{array}{l} B^+_1(\omega)+\alpha \text{e}^{\omega t}x_2, \\ B^-_2(\omega)-\alpha \text{e}^{-\omega  t}x_1,\\B^+_3(\omega_3),  B^-_3(\omega_3),\\ Q, \tilde Q \end{array}$&\hspace{3mm}$0$

\vspace{1mm}\\

\hline\hline
\end{tabular}

The classification results presented in Tables 1-3 are restricted to the systems with non-trivial magnetic fields and non-trivial Pauli terms.
 The complete list of the inequivalent systems with trivial magnetic field can be found in  paper \cite{20_2}.  For the readers convenience and for our future needs we present  the remaining   systems with trivial vector potentials in  Table 4 where the last item is needed only for the more general equation considered in the following sections.

\begin{center}Table 4.
Symmetries   of systems with pure electric field.
\end{center}
\begin{tabular}{l l l l}
\hline No&Potential $A^0$ &Symmetries&Potentials $\tilde A$
\vspace{1mm}

\\

\hline
\vspace{2mm}

1.&\hspace{-3mm}$\begin{array}{l}\frac1{\tilde r^2}R(\varphi)\end{array}$&
 $\begin{array}{l}D, A,
P_3, G_3 \end{array}$&$\ln(\tilde r)\tilde R(\varphi)$\\
2.&\hspace{-3mm}$\begin{array}{l} \frac{\kappa}{\tilde r^2}\end{array}$&
 $\begin{array}{l}D, A, J_3,
P_3,  G_3 \end{array}$&$ {\mu}\ln({\tilde r})$\\

3.&\hspace{-3mm}$\begin{array}{l}
\frac\kappa{x_2^2}\end{array}$ \vspace{1mm} &$
\begin{array}{l}A,  D, P_1,  P_3,\\  G_1, G_3, J_2\end{array}$&$\begin{array}{l}
\mu\ln({x_2})\end{array}$\\
4. $\hspace{0mm} $&\hspace{-6mm}
\vspace{1mm}
$\begin{array}{l} R(r)\end{array}$&$\begin{array}{l}J_1,  J_2,  J_3\end{array}$&$\tilde R(r)$
\\
\vspace{1mm}

5.$\hspace{0mm}$\vspace{1mm}& \hspace{-6mm}$\begin{array}{l} R(x_3)\end{array}$&$\begin{array}{l} P_1, P_2, G_1,  G_2,  J_3 \end{array}$&$\tilde x
R(x_3)$\\

\vspace{1mm}

6. \vspace{1mm} \vspace{1mm}&\hspace{-6mm}$\begin{array}{l}R(x_3)-\frac{\omega^2\tilde r^2}2\end{array}$&$\begin{array}{l} B^+_1(\omega_1),  B^-_1(\omega_1),\\  B^+_2(\omega_2), B^-_2(\omega_2),  J_3 \end{array}$&$\tilde R(x_3)$\\

7.  \vspace{1mm}&\hspace{-8mm}$\begin{array}{l}-\frac{\omega_1^2x_1^2}2 -\frac{\omega_2^2x_2^2}2 \end{array}$&$\begin{array}{l}  B^+_1(\omega_1),  B^-_1(\omega_1), P_3,\\  B^+_2(\omega_2), B^-_2(\omega_2),  G_3 \end{array}$&\ \ 0\\

 8. \vspace{1mm}&\hspace{-7mm}$\begin{array}{l}-\frac{\omega_1^2x_1^2}2-
\frac{\omega_2^2x_2^2}2-\frac{\omega_3^2x_3^2}2\end{array}$
&$\begin{array}{l} B^+_1(\omega_1), B^-_1(\omega_1),\\  B^+_2(\omega_2), B^-_2(\omega_2) ,\\ B^+_3(\omega_3), B^-_3(\omega_3) \end{array}$&\ \ 0\\

9. \hspace{0mm} \vspace{1mm}&\hspace{-6mm}$\begin{array}{l} -\frac{\omega^2x_3^2}2\end{array}$
&$\begin{array}{l} P_1, P_2, G_1, G_2 J_3 \\B^+_3(\omega), \ B^-_3(\omega), \ \end{array}$&\ \ 0\\

 10. \hspace{0mm} \vspace{1mm}&\hspace{-6mm}$\begin{array}{l}-\frac{\omega^2\tilde r^2}2\end{array}$&$\begin{array}{l}  B^+_1(\omega_1),  B^-_1(\omega_1),\\  B^+_2(\omega_2), B^-_2(\omega_2) , \\J_3,  P_3, G_3 \end{array}$&\ \ 0\\

11. \vspace{1mm}&\hspace{-6mm}$\begin{array}{l}-\frac{\omega^2\tilde r^2}2-
\frac{\omega_3^2x_3^2}2\end{array}$
&$\begin{array}{l} B^+_1(\omega_1),  B^-_1(\omega_1),\\  B^+_2(\omega_2), B^-_2(\omega_2) ,\\ B^+_3(\omega_3), B^-_3(\omega_3), J_3 \end{array}$ &\ \ 0\\

12.&\hspace{-4mm} $\begin{array}{l}\frac\kappa {r^{2}}\end{array}$& $\begin{array}{l}A, D,
J_1, J_2,  J_3,\\ {\text { and }} \hat Q {\text { for}}\\{\text { Hamiltonian }} (\ref{H3333})\end{array}
$&$\mu\ln({r})$\\

13. $\hspace{0mm} $&\hspace{-6mm}
\vspace{1mm}
$\begin{array}{l} \frac1{2\nu}\ln r\end{array}$&$\begin{array}{l}J_1,  J_2, J_3, \\ {\text { and }} \hat Q {\text { for}}\\{\text { Hamiltonian }} (\ref{H3333})\end{array}$&\hspace{3mm}0 \\

\hline\hline
\end{tabular}

\vspace{2mm}

Thus we complete the classification of symmetries of SP equations with time independent potentials. Such equations keep all symmetries of the SEs with scalar and vector potentials (which, however, should be slate
modified by changing orbital momenta by the total orbital momenta) and can admit additional matrix symmetries.  Notice that the analogous result has been proven earlier for the stationary Schr\"odinger equation  \cite{maga}.

\section{Quasi relativistic Schr\"odinger equation}

It is pretty well known that SP equation is nothing but the first approximation of the relativistic Dirac equation provided the approximation parameter is the inverse speed of light $\frac1c$. Moreover, the Dirac equation predicts the fixed and correct value of the coupling constant $g$ in equation  (\ref{H1}).

Considering the next approximation (which can be found starting with the Dirac equation and applying the Foldy-Wouthuysen  transformation in the approximate \cite{Foldy} or exact \cite{N21} forms) we come to the quasi relativistic Schr\"odinger equation (QRSE), which, in addition to Pauli term, includes two additional terms describing the spin-orbit and Darwin couplings. The corresponding Hamiltonian being written in vector notations looks as follows:
\begin{gather}\la{H33}H=\frac{{\mathbf \pi}^2}{2m}+neA^0-\frac{ne\hbar}{2mc}{\mbox{\boldmath$\sigma$\unboldmath}}\cdot{\bf  H}-\frac{ne \hbar}{8m^2c^2}{\mbox{\boldmath$\sigma$\unboldmath}}\cdot({\bf E}\times{\mbox{\boldmath$\pi$\unboldmath}}-{\mbox{\boldmath$\pi$\unboldmath}}\times{\bf E})+\frac{ne\hbar}{8m^2c^2}div{\bf E}\end{gather}
where $m$ is a particle mass, $e$ is the electron charge, $\hbar$ is Plank constant, $c$ is the speed of light,  ${\bf E}=\nabla A^0$ and $\pi^a=p^a-\frac{ne}c A^a$.

Notice that in atomic units $\hbar=e=1, c=137$. In this units the electronic mass is normalized to the unity also. But we keep the mass term arbitrary to reserve the possibility to describe a motion of a particle whose mass differs from the electron one. In addition, the described particle can have an  arbitrary charge $ne$ with a positive integer $n$.

In (\ref{H33}) we also keep the terms proportional to $\frac{n^2e^2}{c^3}$ which can be essential for large $n$. For $n=1$ it is reasonable to omit them and reduce the considered Hamiltonian to the following form:
\begin{gather}\la{H333}H=\frac{{\mathbf p}^2}{2m}+eA^0-\frac{e\hbar}{2mc}{\mbox{\boldmath$\sigma$\unboldmath}}\cdot{\bf  H}-\frac{e \hbar}{8m^2c^2}{\mbox{\boldmath$\sigma$\unboldmath}}\cdot({\bf E}\times{\bf p}-{\bf p}\times{\bf E})+\frac{e\hbar}{8m^2c^2}div{\bf E}.\end{gather}

The first three terms in the r.h.s. of equation (\ref{H333})  represent the SP Hamiltonian for neutral particles whose symmetries were classified in recent paper \cite{20_1}. Surely this classification can be extended to the case of systems with spin-orbit couplings, but we will not do it here since  it would need an enormous extension of the paper size.

We will consider also a more general  quasi relativistic Hamiltonian which includes the additional potential $S({\bf x})$:
\begin{gather}\la{H3333}H=\frac{{\mathbf \pi}^2}{2m}+ A^0-\frac{\hbar}{2mc}{\bf{ \sigma}}\cdot{\bf  H}-\frac{\hbar}{8m^2c^2}{\mbox{\boldmath$\sigma$\unboldmath}}\cdot({\tilde{\bf E}}\times{\mbox{\boldmath$\pi$\unboldmath}}-{\mbox{\boldmath$\pi$\unboldmath}}\times{\tilde{\bf E}})+\frac{\hbar}{8m^2c^2}div{\bf{\tilde E}}\end{gather}
where
\begin{gather}\la{efi} \tilde {\bf E}=\nabla \tilde  A, \quad \tilde A=neA^0-qS({\bf x}).\end{gather}

In contrast with (\ref{H33}) the electric field is replaced by the generalised effective vector field $\tilde E$ which is the gradient of the another potential $\tilde  A$. In other words now we have  the additional scalar potential which has clear relativistic roots. Namely, just Hamiltonian (\ref{H3333}) corresponds to the quasi relativistic approximation of the Dirac equation which includes the superposition of the four-vector and scalar potentials. Of course this Hamiltonian is reduced to (\ref{H33}) in the particular case when the coupling constant $q$ is trivial.

We will search for symmetries of QRSE which includes  the most general  Hamiltonian (\ref{H3333}) and  its reduced version (\ref{H33}). First
let us  rewrite (\ref{H33}) in a more simple form using tensor notations and renormalized independent variables and potentials:
\begin{gather} \la{H3} H=\frac12\pi_a\pi_a+A^0+g \sigma_a H^a+\frac\nu2\varepsilon_{abc}\sigma_a(E^bp^c+p^cE^b)+\mu \Delta A^0\end{gather}
where $\Delta$ is the Laplace operator, $\mu$ and $\nu$ are the coupling constants which are proportional to the squared inverse speed of light $\frac1{c^2}$ and $E^b$ are components of the electric strength vector which in our case are proportional  to $\p_bA^0$. The constants $\mu$ and $\nu$ can be treated as rescaled multipliers presented (\ref{H33}), but we will not fix them and treat as arbitrary ones, which can appear as an effect of  the anomalous interaction. Notice that Hamiltonian (\ref{H333}) also can be rewritten in the form analogous to (\ref{H3}) but with $\pi^a $ changed by $p^a$.

The more general Hamiltonian (\ref{H3333}) also can be written in tensor notations:
\begin{gather} \la{H3a} H=\frac12\pi_a\pi_a+\tilde A+g \sigma_a H^a+\frac\nu2\varepsilon_{abc}\sigma_a(\tilde E^b\pi^c+\pi^c\tilde E^b)+\mu (\p_a \tilde E_a)\end{gather}
with potentials and components of the electric field are defined in (\ref{efi}).

The SEs with
Hamiltonians (\ref{H3}) and (\ref{H3a}) are invariant  with respect to neither Poincar\'e nor Galilei group. At the best of my knowledge symmetry of this equation with respect to Lie groups  is still terra incognita. In the following sections we will search for these symmetries.

\subsection{Generalized determining equations}
In Section 2 we have solved the subproblem of our classification problem which corresponds to the trivial coupling constants $\lambda$ and $\nu$, which is a particular but very important part of our task. In the present section we consider  the general case with non-trivial spin-orbit and Darwin terms.

The specificity of the considered classification problems is that the free elements  $\lambda$ and $\nu$ are not dimensionless but should be proportional to the squared inverse  speed (of light), and so it can be changed under the dilatation transformations. However in the atomic units the speed of light is reduced to the dimensionless constant c=137 and we shall treat it as a non changing unit which commutes with symmetry group generators.

Substituting Hamiltonian (\ref{H3}) into the operator equation (\ref{ic}) we again obtain the determining equations (\ref{de1})-(\ref{de6}). However, the remaining equations (\ref{de7}), (\ref{de8}) will include the additional terms proportional to $\lambda$ and $\nu$ which can be indicated in the following formulae:
\begin{gather}\la{24}\begin{split}& \dot\xi^a+\frac12\eta_a
=e(A^b\xi^a_b-\xi^bA^a_b+\alpha A^a)\\&+\frac\nu2(\varepsilon_{bcd}\sigma^cA^0_d\xi^a_b-\varepsilon_{acd}\sigma^c(\xi^kA^0_{kd}+\alpha A^0_d)-\frac12\varepsilon_{acd}[\sigma^cA^0_d,\eta]),\end{split}\\
\label{25}\xi^aV_a=\alpha V+\dot\eta+\ri[\eta,V]-A^a\eta_a-\frac\nu2\varepsilon_{acd}\{\sigma^cA^0_d,\eta_a\}\end{gather}
where $V$ is the modified matrix potential
\begin{gather}\la{26}\begin{split}&V=V^0+g\sigma_a\tilde H^a,\quad \tilde H^a=H^a-2\nu\varepsilon^{abc}A^bE^c, \\&
V^0=A^0+\frac12((A^1)^2+(A^2)^2+(A^3)^2)+\mu \Delta A^0.\end{split}\end{gather}
.

By construction equations (\ref{24}) and (\ref{25}) include $2\times2$ matrices, i.e., in fact any of them is nothing but a system of four equations. Collecting in (\ref{24}) the terms proportional to the unit matrix we again come to equation (\ref{con2}). The terms proportional to matrix $\sigma^c$ look as follows:
\begin{gather}\eta^c_a=\nu(\varepsilon_{bcd}A^0_d\xi^a_b-\varepsilon_{acd}(\xi^kA^0_{kd}+\alpha A^0_d)+2\delta^{ac}\eta^kA^0_k-2\eta^aA^0_c)\la{27}\end{gather} where we again use the expansion (\ref{eta}). In contrast with (\ref{con1}) the r.h.s. of equation (\ref{27}) is not trivial.

Equation (\ref{25}) also can be decoupled to the scalar and vector parts:
\begin{gather}\label{28}\xi^aA^0_a=\alpha V^0+\dot\eta^0-A^a\eta^0_a-\nu\varepsilon_{acd}\eta^c_aA^0_d\end{gather} and
\begin{gather}\label{29}g(\xi^a\tilde H_a^b-\alpha \tilde H^b+2\varepsilon^{bcd}\eta^c\tilde H^d)-\dot\eta^b+\nu\varepsilon_{abd}\eta^0_aA^0_d - \eta^b_aA^a=0.\end{gather}

Substituting (\ref{con2}) into (\ref{28})  and using definition (\ref{26})  we reduce the latter equation to the following form:
\begin{gather}\la{30}\xi^a(A^0_a+\lambda\Delta A^0_a)+\dot\xi^aA^a=\alpha (A^0+\lambda\Delta A^0)+\dot \eta^0-\nu\varepsilon_{acd}\eta^c_aA^0_d. \end{gather}

Finally, substituting the expression (\ref{27}) for $\eta^a_c$ into (\ref{30}) we obtain:
\begin{gather}\la{31}\begin{split}&\hspace{12mm}\xi^a(A^0_a+\lambda\Delta A^0_a)+\dot\xi^aA^a=\alpha( A^0+\lambda\Delta A^0)+\dot \eta^0\\&+\nu^2(2\alpha A^0_dA^0_d+2\xi^kA^0_{kd}A^0_d+\xi^n_nA^0_mA^0_m-\xi^m_nA^0_mA^0_n). \end{split} \end{gather}

Notice that for Hamiltonian (\ref{H3a}) equations  (\ref{27}), (\ref{29}) and (\ref{31}) are generalised to the following form:
\begin{gather}\eta^c_a=\nu(\varepsilon_{bcd}\tilde A_d\xi^a_b-\varepsilon_{acd}(\xi^k\tilde  A^k_d+\kappa\alpha \tilde A_d)+2\delta^{ac}\eta^k\tilde A_k-2\eta^a\tilde A_c),\la{27a}\\\label{29a}g(\xi^a\tilde H_a^b-\alpha \tilde H^b+2\varepsilon^{bcd}\eta^c\tilde H^d)-\dot\eta^b+\nu\varepsilon_{abd}\eta^0_aA^0_d - \eta^b_aA^a=0\end{gather} and
\begin{gather}\la{30a}\xi^a(\tilde A_a+\lambda\Delta \tilde A_a)+\dot\xi^aA^a=\alpha (\tilde A+\lambda\Delta A^0)+\dot \eta^0-\nu\varepsilon_{acd}\eta^c_a\tilde A_d \end{gather}
where $\tilde H^a=H^a-2\nu\varepsilon^{abc}A^b\tilde A_c .$

Thus to describe Lie symmetries of the QRSE with Hamiltonian (\ref{H3}) we are supposed to solve the system of determining equations  (\ref{de1})-(\ref{de6}), (\ref{con2}), (\ref{27}), (\ref{29}) and (\ref{31}), while for Hamiltonian (\ref{H3a}) we have the system (\ref{de1})-(\ref{de6}), (\ref{con2}), (\ref{27a}), (\ref{29a}) and (\ref{30a}). In spite of its frightening view these systems can be effectively solved, since the solution of the main part of them have been obtained in the previous section.

\subsection{Equivalence relations}

Any search for Lie symmetries of a partial differential equation should include a consistent definition of admissible (equivalence) transformations which keep the generic form of this equation but can change its arbitrary elements (in our case potentials $A^0$,
$\tilde A$ and $A^a)$.

Till now we restricted ourselves to the direct  use the equivalence gruppoid for the Schr\"odinger equation found in paper \cite{20_2} since in fact all inequivalent potentials are fixed  by the determining equations  (\ref{de1}), (\ref{de2}), (\ref{de6}), (\ref{con2}) and (\ref{con3}) which are the same as in the case of trivial Pauli, spin-orbit and Darwin terms. A particular sector  of the equivalence relations is formed by gauge transformations which change vector-potentials but keep the vectors of electric and magnetic fields invariant.

 But whenever the spin-orbit coupling is nontrivial, we can find a new effect the gauge transforms which consists in changing the magnetic field present in the Pauli term. Indeed, changing $\psi\to\exp(i \Phi )\psi$ in the generalised Schr\"odinger equation including Hamiltonian  (\ref{H3}) we observe the following change of the Pauli term:
 \begin{gather}\la{wau} g\sigma^aH^a\to g\sigma^a H'^a\end{gather}
 where
 \begin{gather}\la{wauu} H'^a=H^a-\frac\nu{g} \varepsilon^{abc}\Phi_bE^c.\end{gather}

 In particular, for some class of the external electric and magnetic fields the related Pauli term can be reduced to zero. The necessary property satisfied by such fields is that $\bf E$ and $\bf H$ are orthogonal, i.e., $E^aH^a=0$. In other words, one of the invariants of the electromagnetic field should be trivial.

Thus the gauge transformations being applied to the QRSE with Hamiltonians (\ref{H3}), (\ref{H33}) and (\ref{H333}) can change the effective magnetic field, and this effect is rather specific and cannot be observed in Schr\"odinger, Schr\"odinger-Pauli and Dirac equations.
 Important examples of the magnetic field which can be nullified in Hamiltonian (\ref{H3})  by the charge transformations are the constant magnetic field and the  field generated by the infinite straight current.

\subsection{Solution of the determining equations}

Since we have in hands all inequivalent solutions of equations  (\ref{de1})-(\ref{de6}) and (\ref{con2}) which were found in paper \cite{20_2}, we have a priori information about inequivalent versions of vector potentials and symmetries which are represented in Tables 1--3.  Thus the only thing we need is to verify whether these vector potentials are compatible with equations (\ref{27}), (\ref{29}),  (\ref{31}) for Hamiltonians (\ref{H3}) and equations (\ref{27a}), (\ref{29a}),  (\ref{30a}) for Hamiltonians (\ref{H3a}). In addition it is necessary to study matrix integrals of motion which do not include differentials, since they are not fixed by the subsystem of the determining equations

Like in
Section 2 we consider vector potentials which correspond to inequivalent one dimension symmetry algebras. For symmetry presented in (\ref{2_2}) equations determining equations (\ref{con2}), (\ref{27}), (\ref{29}) and (\ref{31}) look as follows:
\begin{gather}\la{con21} \eta^0_a
=-A^a_3,\\\eta^c_a=\nu(-\varepsilon_{acd}A^0_{3d}+2\delta^{ac}\eta^kA^0_k-2\eta^aA^0_c),\la{con22}
\\\label{con23}g(\tilde H_3^b+2\varepsilon^{bcd}\eta^c\tilde H^d)-\dot\eta^b+\nu\varepsilon_{abd}\eta^0_aA^0_d - \eta^b_aA^a=0\end{gather}and
\begin{gather}\la{32}(A^0+\mu\Delta A^0-{\nu^2}A^0_{d}A^0_d)_3=\dot \eta^0. \end{gather}

It was shown in paper \cite{20_2} that up to gauge transformations the generic solution of the scalar equation (\ref{con21}) is given by  equations (\ref{vp1}) and (\ref{2}) while the related function $\dot \eta^0$ is trivial.

Using the identity
\begin{gather}\la{hop}\varepsilon_{abc}(\eta^c_a)_b=0\end{gather} we obtain the following  differential consequence of equations (\ref{con22}):
\begin{gather}\la{33}(\Delta A^0-\nu A^0_aA^0_a)_3=0\end{gather} which is compatible with (\ref{32}) (where $\dot \eta^0$ is trivial ) only for the case $A^0_3=0$. The corresponding functions $\eta^k$ solving equations (\ref{con22}) and  (\ref{con23}) are trivial also, and we again come to the generating functions and symmetries fixed in Item 1 of table 1.

We see that if the SP equation (\ref{H1}), (\ref{sp}) admits symmetry $P_3$ then this symmetry is admitted by the more general equation (\ref{H3}) with non-trivial parameters $\nu$ and $\lambda$. Moreover, this symmetry cannot be generalized by adding a matrix term, i.e., a more general motion constant (\ref{2_2}) can be  present only with trivial functions $\eta^\mu$.

The next system we consider is given by relations (\ref{3}). The corresponding symmetry (\ref{3_1}) again has to be generalised by adding matrix $\eta$ and represented in the form (\ref{3_2}). Moreover,  the  components $\eta^a$ should satisfy condition (\ref{27}). It is possible to verify by direct calculation that the related equations  (\ref{27}), (\ref{29}) and (\ref{30}) are satisfied provided $\eta^1=\eta^2=0, \eta^3=\frac12$. We will not reproduce these calculations since this result in fact is evident: the spin-orbit term is a scalar with respect to rotations since the related electric field is a vector as like as ${\mbox{\boldmath$\sigma$\unboldmath}}=(\sigma^1,\sigma^2,\sigma^3)$ and ${\bf p}=(p^1,p^2,p^3)$.

Consider now the systems whose scalar parts are given by equation (\ref{4}). The related generator (\ref{4_1}) should be generalized to the following form:
\begin{gather}\la{41}
Q=L_3+p_3+\eta^0+\sigma_a\eta^a\end{gather}
and so the corresponding determining equations (\ref{con2}), (\ref{27}), (\ref{28}) and (\ref{29}) look as follows:
\begin{gather}\eta^0_a=\varepsilon^{3ab}A^b-A^a_\kappa,\la{42}\\
\eta^c_a=\nu(\delta^{3c}A^0_a-\delta^{ac}A^0_3-\varepsilon_{acd}A^0_{d\kappa}
+2\delta^{ac}\eta^kA^0_k-2\eta^aA^0_c)\la{43},\\
\label{44}g(\tilde H_\kappa^b+2\varepsilon^{bcd}\eta^c\tilde H^d)-\dot\eta^b+\nu\varepsilon_{abd}\eta^0_aA^0_d - \eta^b_aA^a=0,\\\la{45}(A^0+\lambda\Delta A^0-{\nu^2}A^0_{d}A^0_d)_\kappa=\dot\eta^0\end{gather}
where $A^0_\kappa=\p_\kappa A^0,\ \tilde H^b_\kappa=\p_\kappa \tilde H^b,$ and $\kappa=\arctan\frac{x_2}{x_1}+x_3.$

It is possible to show that up to gauge transformations the generic solution of equation (\ref{42}) can be presented in the form given  in Item 3 of Table 1, see paper \cite{20_2}. The related function $\dot \eta^0$ is trivial.

The differential consequence (\ref{hop}) for vector equations (\ref{43}) takes the following form:
 \begin{gather}\la{61}(\Delta A^0-\nu A^0_aA^0_a)_\kappa=0\end{gather} which is compatible with (\ref{44}) only for the case $A^0_\kappa=0$. The corresponding functions $\eta^k$ solving equations (\ref{con22}) and  (\ref{con23}) are trivial, and we  came to the generating functions and symmetries fixed in Item 3 of Table 1.

 The symmetries discussed in the above are valid for both Schr\"odinger-Pauli and quasirelativistic Hamiltonians. In contrary, symmetries of the  system specified  by relations (\ref{5})  are valid for SP equation but not  for QRSE with Hamiltonian (\ref{H3}) .

  To prove this facts it is sufficient to consider the related determining equations  (\ref{con2} ), (\ref{27}), (\ref{29}) and (\ref{31}) which  are reduced to the following form:
 \begin{gather}\la{101}\begin{split}&\eta^0_1=A^1+\mu(A^1+2A^1_y),\ \  \eta^0_2=A^2-\mu(A^1+2A^2_y),\quad \eta^0_3=A^3,\end{split}\\\la{102}\eta^c_a=\nu(\varepsilon_{acd}((1-\kappa)A^0_d- A^0_{dy})
 -2\eta^aA^0_c+2\delta^{ac}\eta^bA^0_b+\mu\delta^{3c}A^0_a),\\\la{103}g(\tilde H^b_y-\alpha \tilde H^b+2\varepsilon^{bcd}\eta^c\tilde H^d)-\dot\eta^b+\nu\varepsilon^{abd}\eta^0_aA^0_d=0,\\
 \la{104}(1+\mu\Delta)A^0_y=\alpha(1+\kappa\Delta)A^0+\dot\eta^0+
 \nu^2(2(1+\kappa\alpha)A^0_dA^0_d+(A^0_dA^0_d)_y)\end{gather}
 where $y=\mu\ln \tilde r+\varphi$.

 In accordance with \cite{20_2} the generic solution of equations (\ref{101}) can be represented in the form given in Item 4 of Table 1, and the related function $\dot\eta^0$ present in equation (\ref{104}) is trivial. Considering again the differential consequence (\ref{hop}) of equation (\ref{102}) and comparing it with (\ref{104})  conclude that the system (\ref{101})--(\ref{104}) is incompatible.

Considering in the analogous manner the potentials and symmetry  fixed in (\ref{6})  and (\ref{6_1}) we come to the following version of the determining equations  (\ref{con2}), (\ref{27}), (\ref{29}) and (\ref{31}):
\begin{gather}\la{con21a} \eta^0_a=-\exp(\omega t)(\omega\delta_{a3}+A^a_3)
\\\eta^c_a=\nu(-\varepsilon_{acd}A^0_{3d}\exp(\omega t)+2\delta^{ac}\eta^kA^0_k-2\eta^aA^0_c)\la{con22a}
\\\label{con23a}g(\exp(\omega t)\tilde H_3^b+2\varepsilon^{bcd}\eta^c\tilde H^d)-\dot\eta^b+\nu\varepsilon_{abd}\eta^0_aA^0_d - \eta^b_aA^a=0\end{gather}and
\begin{gather}\la{32a}(A^0+\mu\Delta A^0-{\nu^2}A^0_{d}A^0_d)_3=\dot \eta^0. \end{gather}

It was shown in  \cite{20_2} that up to gauge transformations the generic solution of (\ref{con21a}) is given by  equation (\ref{2}) while the related function $\dot \eta^0$ is equal to $\omega x_3$. Considering  the differential consequence (\ref{hop}) of relations (\ref{con22a})  and comparing it with (\ref{32a}) we conclude that these equations are incompatible, and so symmetry (\ref{6_1}) generalised by adding a matrix $\eta$ is not accepted by the QRSE  with Hamiltonian (\ref{H33}). The same is true for symmetry (\ref{7_1}), and to prove this fact it is sufficient to repeat the above manipulations staring with equation (\ref{con21a}).

Thus we specify all inequivalent one dimensional symmetry algebras which can be admitted by the QRSE with Hamiltonian  (\ref{H3}). It happens that symmetries (\ref{2_2}), (\ref{3_2}) and (\ref{41}) are valid provided the potentials have the form presented in Items 1, 2 and 3 of Table 1. The same is true for linear combinations of these symmetries specified in the remaining items of this table. In contrary, the symmetries  (\ref{5_1}), (\ref{6_1}) and (\ref{7_1}) are not valid.

 \subsection {Matrix integral of motion}
 The main result of the previous section can be formulated as follows:  Lie symmetries of the SP equation presented in Table 1 are kept also for the QRSE. The only exceptions are the matrix symmetries (\ref{23})  the first of which cannot be kept for any spin-orbit coupling term while the second one is possible if the scalar potential satisfies some additional restrictions.

 In this section we find matrix symmetries for the QRSE. The generic form of the corresponding symmetry operators again is given by formulae (\ref{so}) and (\ref{eta}) where, however, functions $\xi^0$ and $\xi^a$ are trivial. In this case the determining equations  (\ref{con2}), (\ref{27}), (\ref{29}) and (\ref{31}) are reduced to the following ones:
 \begin{gather}\la{105} \eta^0_a=0,\quad \dot\eta^0=0,\\
 \la{107}\dot\eta^a=2g(x_bA^b)_a-(2g+2g\nu+1)A^a+\frac{x_ax_bA^b}{r^2}-x_bA^a_b,\\\la{108}\eta^c_a=
 2\nu(\delta^{ac}\eta^bA^0_b-2g\eta^aA^0_c).\end{gather}

 Equation (\ref{108}) is solved by the following functions:
 \begin{gather}\la{109}A^0=\frac1{2\nu}\ln(r),\quad \eta^a=\phi\frac{x_a}r\end{gather}
 where $\phi$ is a function of $t$ which is reduced to a constant if equation (\ref{107}) is satisfied. .

 Thus we discover the following integral of motion:
 \begin{gather}\la{110}\hat Q=\frac{\sigma^ax_a}r\end{gather} which commute with Hamiltonian (\ref{H3}) provided scalar potential $A^0$ is given by equation (\ref{109}) and vector potential $A^a$ satisfies condition (\ref{107}) with $\eta^a$ given by equation (\ref{109}). The generic form of the vector-potential satisfying this condition is given by the following equation:
 \begin{gather}\la{111}A^a=\frac{\varepsilon^{abc}\phi^bx_c}{r^{1+\nu+\frac1g}}, \ g\neq0 \end{gather}
 where $\phi^b, b=1,2,3$ are arbitrary functions of angular variables $\varphi$ and $\theta$.

  In particular vector potential can be trivial, then the corresponding QRSE with scalar potential (\ref{109}) admits symmetry (\ref{110}) and three additional symmetries \begin{gather}\la{112}J^1=L^1+\frac12\sigma_1,\quad J^2=L^2+\frac12\sigma_2, \quad J^3=L^3+\frac12\sigma_3\end{gather} where
$L_1, L_2$ and $L_3$ are components of the orbital momentum given by equation (\ref{gen}). For nontrivial  vector potentials represented in (\ref{112}) this set of symmetries is missing. However  if functions $\phi^a$ are constant there is an additional integral of motion, namely, a linear combination of $J^a$. In this case up to rotation transformation we can set $\xi^1=\xi^2=0, \xi^3=const$, then  the additional integral of motion is reduced to $J^3$.
\section{Summary on symmetries of QRSE}

 In the previous sections we specify Lie symmetries of the SP equation which are valid also for the quasi relativistic Schr\"odinder equation. In particular, the complete list of potentials generating non-equivalent symmetries for QRSE  with Hamiltonian (\ref{H3})
 have been obtained and fixed in Table 1. Notice that the symmetries presented in this table are valid also for the SP equation, and the addition spin-orbit term presented in the generalized Hamiltonian  (\ref{H3})  keeps all of them. However, the symmetries of SP equation presented in Tables 2--4 are lost provided the spin-orbit coupling be introduced.

 Let us stress that the more general QRSE with Hamiltonian (\ref{H3a}) keeps all symmetries of the SP equation provided the potential $\tilde A$ has some special forms. Moreover, for the special potentials presented in Items 12 and 13 of Table 4 we have the additional matrix symmetries which are missed in the case of SP equation. To  prove these facts it is sufficient to repeat the calculations presented in Section 3.3 for a bit more general system of the determining equations (\ref{27a}), (\ref{29a})  and (\ref{30a}) which are rather similar to (\ref{27}), (\ref{29}) and (\ref{31}) but includes the additional arbitrary element $\tilde A$. We will not present here the  calculation details but restrict ourselves to the presentation of their results in the right columns of Tables 2--4.

 \section{Discussion}

We present the completed group classification of Schr\"odinger-Pauli and the quasi relativistic SEs. In other words we create  certain group-theoretical grounds for all quasirelativistic quantum mechanical models of the spin-orbit and Pauli couplings. The main result of our research consists in obtaining the  a priori information about all inequivalent Lie symmetries which can be accepted by such models. Moreover, this information is constructive since all inequivalent potentials of the external electromagnetic field involved into such models are specified also.

The presented results can be treated as a direct extension of the basic symmetries of the free Schr\"odinger equation \cite{Nied, And} to the case of more general equations including potentials and the above mentioned coupling terms. In some sense these results are complete since we consider almost all possible relativistic correction of order $1/c^2$, and it is well known that the corrections of higher order in the inverse speed of light are not reasonable for the one particle SEs. The only term of order $1/c^2$ which we ignore is the relativistic correction to the kinetic energy $p^4/8m^3$. Its presence reduces the number of symmetries admitted by the quasi relativistic equation, but not too dramatically. Namely,  symmetries presented in Table 1 are kept  if this  term is added, but the generators of pure Galilei transformation $G_2$ and $G_3$ should be excluded. The same is true for the systems fixed in Items 1--3 of Table 4. However, symmetries presented in Tables 2 and 3 and Items 4--13 of Table 4 are not valid in presence of the relativistic correction of the kinetic energy.

Notice that symmetries of the SEs with the relativistic correction of the kinetic energy were discussed in old paper \cite{FNFN}.

Let us remind that the Pauli term in Hamiltonians (\ref{H2}) and (\ref{H3}) is not necessary to be interpreted as a relativistic correction, since it is naturally generated  (with the correct coupling constant) by the Galilei invariant Levi-Leblond equation \cite{LL}. Moreover, the spin-orbit and Darwin terms in  (\ref{H3}) also can be obtained starting with first order Galilei invariant motion equations \cite{N20}. But in contrast with the Dirac equation, Galilei invariant ones do not predict a fixed value of the related coupling constants.

Like in papers \cite{Nuca, 20_2} to solve our classification problem we apply the algebraic approach whose main idea is the systematic use of the subgroup structure of the equivalence group. An important element of our analysis was the fixation of the equivalence qroupoid of  the analysed class of equations. It happens that this gruppoid has the same structure as in the case of the standard Schr\"odinger equation with arbitrary scalar and vector potentials, thus we could simple use the results of paper \cite{20_2} which form a ground for the study of more general equations.  We do not present the equivalence relations here since for the  sets of potentials presented in Tables 1-4 these relations are fixed in \cite{20_2}. The only potentials missing in \cite{20_2} are given by equations (\ref{108}) and (\ref{109}), but the corresponding  equivalence relations belong to the symmetry group of the QRSE or are realized by the gage transformations.

 The latter potentials correspond to the system with a rather exotic integral of motion being a matrix dependent on spatial variables which is fixed by equation (\ref{110}). Notice that in the case when the vector potential $A^a$ is trivial this system appears to be maximally superintegrable, since, in additon to the symmetries indicated in Item 2 of Table 4  there exist one more matrix integral of motion
 \begin{gather}\la{112a}\tilde Q=\sigma^aL^a+1.\end{gather}

 Operators (\ref{110}) and (\ref{112a}) commute with the Hamiltonian and total orbital momentum, but anticommute  between themselves. Thus we have a perfect example of symmetry superalgebra whose odd elements are $\tilde Q$ and $ \hat Q$ while the even elements are $J^1, J^2, J^3, J^aJ^a$ and the unity operator.

 Alternatively, we can apply one more integral of motion
   \begin{gather}\la{113} Q=(\sigma^aL^a+1)P\end{gather} where $P$ is the parity operator. Operators (\ref{110}) and (\ref{113}) commute each other, and so we can indicate four commuting integrals of motion,  say, $J^3,\ J^aJ^a, \ \hat Q $ and  $Q$ which make our system with three spatial and one spin  degree of freedom integrable. The remaining integrals of motion guarantee the superintegrability of the considered system. For the other examples of the integrals of motion for the SP equations including the parity and some other discrete symmetries see papers  \cite{N3} and \cite{N4}.

   Notice that up to the matrix multiplier $\gamma^0$ operator $\tilde Q$ coincides with  the "Dirac constant of motion" for the Dirac equation with spherically symmetric potential of the external electric field. In equation (\ref{113}) matrix $\gamma^0$ is changed by the parity operator, and in this form $Q$ is valid for the Dirac equation also provided $P$ is the Dirac parity operator.

   The considered models and the generic discussion presented above  looks as a good stimulation to study superintegrable systems which include both the Pauli and spin-orbit couplings. Indeed, the correct QRSE should include both of them. We remind that the superintegrability aspects of  systems with spin-orbit coupling (but without Pauli one) and Pauli coupling (but without spin-orbit one) are relatively well studied, refer to \cite{WY1, WY2,WY3} and \cite{N5, N6, N7}. Thus the next step is to investigate the systems including both these couplings.

   One more challenge is to classify symmetries of the QRSE with Hamiltonian (\ref{H333}). In the case when the external electric field is trivial this equation is reduced to the SP equation for neutral particles whose symmetries are classified in recent paper \cite{N4}.

\vspace{2mm}

{\bf Acknowledgement}

\vspace{1mm}

The author is indebted to the National Academy of Science of the Ukraine for the financial support in frames of the Programm of Support of the Priority Research, project  7/1/241.

\end{document}